\magnification = 1000
\def\nh{\noindent\hangindent=1 true cm \hangafter = 1}
\baselineskip = 16pt
\tolerance = 10000   
\hsize = 4.25 in
\hoffset = .85 in

\def\b{\noindent}
\def\u{\vskip  .1 in}

The use of cortical field potentials rather than the details of
spike trains as the basis for  cognitive information processing is
proposed. This results in a space of cognitive elements with
natural metrics. Sets of spike trains may also be considered to
be points in a multidimensional metric space. The closeness 
of sets of spike trains in such a space implies 
the closeness of points in the resulting function space of potential
distributions.

\u
\b {\bf 1. Introduction}  
\u

\b Nearly all theories of information processing  focus 
attention on dynamical  patterns of action potentials [1]
including studies which involve
correlational analyses [2].
 In many of these
studies authors choose to regard spikes as temporally discrete 
events and consider their rates or temporal relationships to
be significant. (In reality action potentials are continuous, at
least
within the framework of classical physics.) The usual approach leads to 
difficulties in the construction of metrics for cortical
activity because  metrics for sequences (of time points)
give large distances if minor differences occur between 
spike trains. The latter seems to imply that spike trains themselves
are not a useful description per se in the description of cortical
activity.

\u
\b {\bf 2. Potential distributions and information processing}
\u

\b Rather than focus on spikes and in particular their times of
occurrence as  a point process, we concentrate on
the {\it field potentials} they are associated with, or generate,
in a region   $M \subset R^3$ of cerebral cortex or other brain
structure. It is also convenient to restrict attention to
a finite time  
interval $T = [t_1,t_2] \subset R^+$, say. 

\u
\b Firstly we consider the actual electrical potential distribution
$V({\bf x},t), {\bf x} \in M, t \in T$, throughout the region.
For convenience only extracellular points of $M$ will be considered,
but we will continue to use the same symbol $M$ for the spatial region.
The exact potential $V$ cannot be measured exactly but is able to
be estimated by an approximate {\it field potential}
[3] 
 over a region
$A_{\bf x}$ surrounding the space point ${\bf x}$, 
 (and probably
also a small time interval surrounding $t$)
$$ V_E ({\bf x},t) = {1 \over |A_{\bf x}|}\int\int\int_{A_{\bf x}}
 V({\bf x'},t) w({\bf x'},t) d{\bf x'},
$$
where $|.|$ denotes volume. The region $A_{\bf x}$ reflects the size
of a recording electrode and the weight function $w({\bf x},t)$ reflects
its electrical properties.

\u
\b Now the set of all possible potential
distributions  $V({\bf x},t)$ on $M \times T$,
 which we denote by $C_V(M \times T)$, is a set of bounded continuous
functions and a subset of the space $C(M \times T)$ of all bounded
continuous functions on that product space. For such a space there
are suitable metrics. Let $V_1$ and $V_2$ be two points in $C_V$
( i.e., potential distributions). Then one metric (distance function)
is provided by the uniform or sup norm,
 
$$ d_1(V_1,V_2)= \sup_{M \times T} |V_1({\bf x},t) - V_2({\bf x},t)|. $$
Alternatively and perhaps more satisfactorily we may consider
$C_V$ as a subset of the space of square integrable functions
on $M \times T$ in which case we may use the metric 
$$ d_2(V_1,V_2)= \bigg[\int_M\int_T\big(V_1({\bf x},t) - V_2({\bf x},t)
\big)^2 dtd{\bf x}\bigg]^{1\over 2}. $$
\b However, if one is only interested in comparing potential distributions 
at a given time $t$ and therefore considered only to be functions of {\bf x}, 
then the following corresponding metrics will be useful: 

$$ D_1(V_1,V_2)= \sup_{M } |V_1({\bf x},t) - V_2({\bf x},t)| $$
and

$$ D_2(V_1,V_2)= \bigg[\int_M\big(V_1({\bf x},t) - V_2({\bf x},t)
\big)^2 d{\bf x}\bigg]^{1\over 2}. $$

\u
\b {\bf 3. Applications}
\u
\b Consider a stimulus $S_0$, of extrinsic or intrinsic origin, or a combination of
both. With this stimulus will be associated a set of spike trains,
constituting a point in a metric space - see below. There will also
be an associated potential distribution which we assume is in the
region $M \times T$. However it is very unlikely that either
the set of spike trains or the potential distribution are uniquely
determined by $S_0$ as the response to the same stimulus is never
exactly the same at different presentations. Thus there will be an
average potential distribution associated with $S_0$ which we denote
by $ \overline{V_0}({\bf x},t), {\bf x} \in M, t \in T $. 
If now a stimulus $S$ occurs,
it will be identified with $S_0$ if the potential distribution elicited
, $V$, satisfies
$$ d_1(V, \overline{V_0}) < \epsilon_1, $$
or
$$ d_2(V, \overline{V_0}) < \epsilon_2, $$ 
where the positive constants $\epsilon_1$ and $\epsilon_2$ are 
measures of the discriminatory ability of cognitive processes.

\u
\b {\it Spike Trains}
\u
\b Suppose there are $n$ neurons in the region $M$ and in response to
the stimulus $S$ let the $k-$th of these have spikes at times
$t_{k,1}, t_{k,2},...,t_{k,n_k}$ where all these time points are in $T$.
Let, for $t \in T$, $N_k(t)$ be the number of spikes of neuron $k$ 
in $(t_1,t]$.  Then $\{ N_k(t), t\in [t_1,t_2]\}$ is, for each $k$,
 a right-continuous function on $T$  and is an element of the space
$D(T)$ of functions which are at each
point in $T$ right-continuous and with left-hand limits ({\it cadlag}).
$D(T)$ is a metric space with the uniform norm. Thus we may consider
the whole set  of action potentials in $M$ in $T$ as a point in the
space $D^n(T) $. Let $y_1$ and $y_2$ be two points in $D^n(T)$. Then the distance
between these two sets of action potentials is
$$\rho (y_1,y_2) = \sum_{k=1}^n \rho_k $$
where $\rho_k$ is the distance between the responses in the $k-$th spike
train (supremum on $T$ for each component). We claim that if stimuli 
$S_1$ and $S_2$ lead to sets of spike trains $y_1$ and $y_2$, then
there will be a $\delta$ such that
$$ \rho (y_1, y_2) < \delta  \Rightarrow d_1(V_1,V_2) < \epsilon_1.$$
That is distinguishable stimuli lead to distinguishable sets of
spike trains which are components of distinguishable potential
distributions. Differences in spike train details are expected
to be smoothed out so that minor differences are not relevant 
for cognitive information processing.
\u
\b Another possibility is that the space $C_V$  is partitioned into a
set of disjoint subsets $C_{\epsilon}$ and that when a potential 
distribution falls within $C_{\epsilon}$, the corrresponding
cognitive element pertains.

\u
\b Support from an INSERM fellowship and an invited professorship at Universit\'e
Paris VI through  R\'emy Lestienne 
  are appreciated.
\u

\b {* \it Email: tuckwell@b3e.jussieu.fr}

\u

\nh  [1]  Abeles,M., H.Bergman, I.Gat, I.Meilijson, E.Seidemann, 
 N.Tishby and E.Vaadia,
   Proc. Natl. Acad. Sci. USA {\bf 92}, 8616 (1995); Kruglyak,L. and
   Bialek,W., Neural Comp. {\bf 5}, 21 (1993); Rodriguez,R. and Tuckwell,H.C.
 Phys. Rev. E {\bf 54}, 5585 (1996);
 Hubel,D.H. \&  Wiesel,T., 
 J. Physiol. {\bf 160}, 106 (1962);
Konig,P., Engel,A.K. \& Singer,W., Trends Neurosci.
 {\bf 19}, 130 (1996);
 Lestienne,R. \&  Tuckwell,H.C. 
 Neuroscience {\bf 82}, 315 (1998);
 Mainen,Z. \&  Sejnowski,T., Science {\bf 268}, 1503 (1995).

\nh  [2] Toyama,K. \&  Tanaka,K.   In, {\it Dynamic Aspects of Cortical Function},
 G.M.Edelman et al., eds. Wiley, New York (1984).
    
\nh  [3]  Bush,P. \&   Sejnowski,T.,  J. Comp. Neurosci. {\bf 3}, 91 (1996);
 Livingstone,M.S.,  J. Neurophysiol. {\bf 75}, 2467 (1996).

\vfill\eject\end